\documentstyle[11pt,newpasp2,twoside,epsf, psfig, epsfig]{article}
\def\arg#1{{\it#1\/}}

\newbox\grsign \setbox\grsign=\hbox{$>$}
\newdimen\grdimen \grdimen=\ht\grsign
\newbox\laxbox \newbox\gaxbox
\setbox\gaxbox=\hbox{\raise.5ex\hbox{$>$}\llap
        {\lower.5ex\hbox{$\sim$}}}\ht1=\grdimen\dp1=0pt
\setbox\laxbox=\hbox{\raise.5ex\hbox{$<$}\llap
        {\lower.5ex\hbox{$\sim$}}}\ht2=\grdimen\dp2=0pt
\newcommand{\simlt}{\mathrel{\copy\laxbox}}
\newcommand{\simgt}{\mathrel{\copy\gaxbox}}

\let\prog=\arg

\def\edcomment#1{\iffalse\marginpar{\raggedright\sl#1\/}\else\relax\fi}
\reversemarginpar

\begin{document}
\title{X-ray properties of spiral galaxies}
 \author{Roberto Soria}
\affil{Mullard Space Science Laboratory, University College London, 
Holmbury St Mary, Surrey RH5 6NT, UK; rs1@mssl.ucl.ac.uk}

\begin{abstract}

X-ray studies of nearby spiral galaxies with star formation 
allow us to investigate temperature and spatial distribution 
of the hot diffuse plasma, and to carry out 
individual and statistical studies of different classes 
of discrete sources (low- and high-mass X-ray binaries, 
Supernova remnants, supersoft and ultra-luminous sources). 
In particular, we briefly review the different models proposed  
to explain the ultra-luminous sources. 
We can then use the X-ray properties of a galaxy 
to probe its star formation history.
We choose the starburst spiral M\,83 
to illustrate some of these issues.

\end{abstract}

\section{Introduction}

It is now thirty years since the first unambiguous 
identification of X-ray emission from our nearest giant 
spiral, M\,31 (Margon, Lampton \& Cruddace 1974).  
Today we know that X-ray emission in galaxies comes 
from discrete sources and hot diffuse gas. 
Discrete sources (accreting compact 
objects and SuperNova Remnants) are a fossil record 
of the stellar population, and may be used as a probe 
of star formation history. 
Diffuse X-ray emission is the tracer of the hot gas ($T> 10^6$ K) 
in galaxies and clusters; in galaxies, it is an indicator of recent 
star formation. Disentangling truly diffuse gas from faint, 
unresolved sources---such as faint X-ray binaries (XRBs), cataclysmic 
variables, coronal emission from main-sequence and T Tauri stars---
is still a major problem, despite the fact that 
sensitivity and spatial resolution of the X-ray 
detectors have improved by three orders 
of magnitude since those early observations.

Today, X-ray studies of nearby galaxies 
can be conducted on two complementary levels.
We can do a statistical study of the spatial and 
spectral distribution of the sources: this can help us   
distinguish between different physical classes 
of X-ray emitters.
And we can use the X-ray sources as a tool to probe 
the structure and evolution of the host galaxy.

Statistical studies of X-ray sources 
in the Milky Way are hampered by the large 
relative uncertainty in the distance of most sources 
(often by a factor of two) and by our very incomplete 
view due to the large extinction 
in the Galactic plane. Population studies 
can be conducted more easily 
in nearby galaxies, with more favorable 
viewing angles and the same relative distance 
for all the sources. By studying a large sample 
of galaxies it is also possible to quantify 
how the X-ray properties of a galaxy depend on its structural 
type and level of star-forming activity.
Moreover, by quantifying the relation between 
observed X-ray properties and star formation history 
in nearby galaxies, one may predict the luminosity 
and color distribution of the faint galaxies 
detected in the Chandra Deep Field surveys 
(Giacconi et al. 2002),
and therefore probe star formation at high redshift. 
(See R. Griffiths's contribution 
elsewhere in these Proceedings.)

In this conference paper, I have chosen  
the starburst galaxy M\,83 to illustrate some of these issues. 
Located at a distance of $\approx 4$ Mpc, 
M\,83 is a grand-design spiral seen 
at low inclination.
More than 100 sources are resolved in a 51 ks {\it Chandra} 
observation. In addition to the point sources, 
a true-color image (Fig.\ 1) shows a bright starburst nucleus 
and strong, soft diffuse emission along the spiral arms.


\begin{figure}  
\plotone{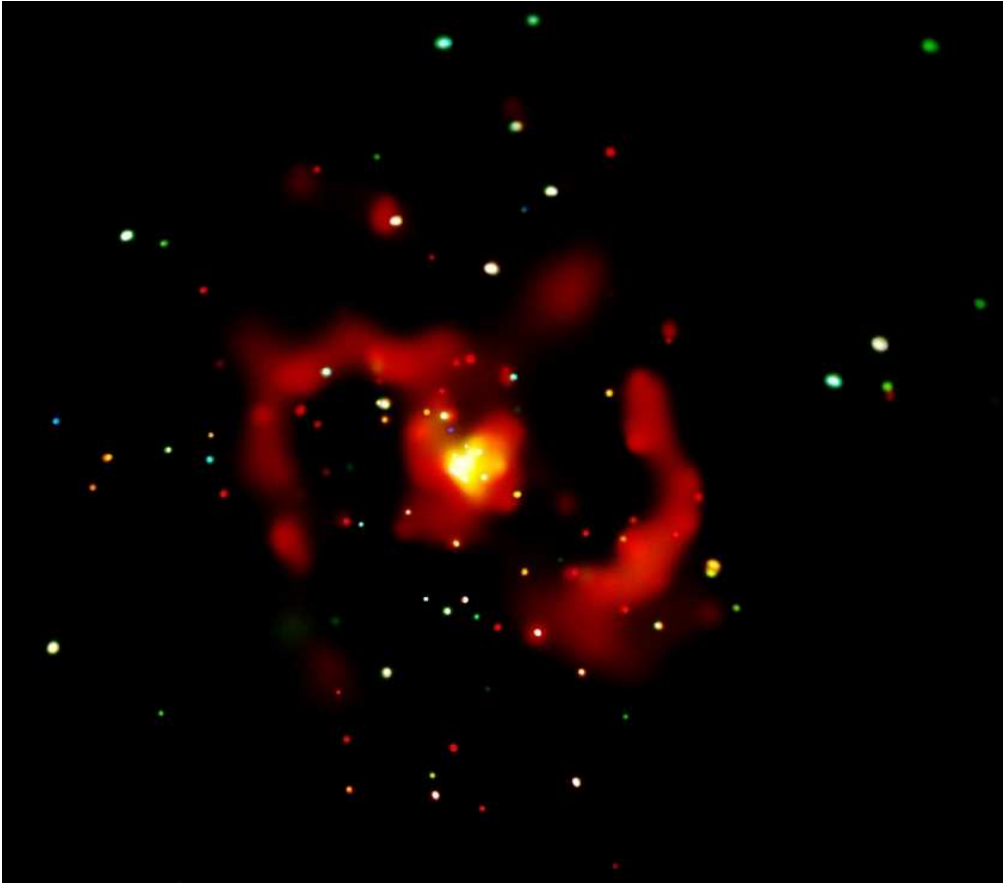}
\caption{{\footnotesize A true-color {\it Chandra}/ACIS image of M\,83 shows 
about 130 discrete sources and diffuse emission 
in the starburst nucleus and along the arms. The colors are: 
red $= 0.3$--1.0 keV; 
green $= 1.0$--2.0 keV; blue $= 2.0$--8.0 keV. Size of the image: 
12\arcmin $\times$ 10\arcmin. North is up, East is left.}}
\end{figure}

\nopagebreak

\section{Diffuse X-ray emission}

The total X-ray luminosity emitted from M\,83 in the 
0.3--8.0 keV band is $\approx 2.5 \times 10^{40}$ erg s$^{-1}$
\footnote{In other bands: $L_{\rm B} \approx 2.5 \times 10^{43}$ erg s$^{-1}$, 
$L_{\rm FIR} \approx 2.5 \times 10^{43}$ erg s$^{-1}$, 
$L_{\rm H\alpha} \approx 1.5 \times 10^{40}$ erg s$^{-1}$. 
In general, $L_{\rm X} \sim L_{\rm H\alpha} \sim 10^{-3} L_{\rm FIR}$ 
for starburst spirals 
(Fabbiano \& Shapley 2002; Calzetti et al. 1995; 
Condon et al. 1998; de Vaucouleurs et al. 1991); all three 
bands can be used as indicators of star formation.}. 
Of this, $\approx 6 \times 10^{39}$ erg s$^{-1}$ comes 
from the resolved sources and the rest is unresolved.
The nuclear starburst contributes for 
$\approx 5 \times 10^{39}$ erg s$^{-1}$ (of which 
$\approx 1 \times 10^{39}$ erg s$^{-1}$ from resolved sources).
The unresolved X-ray emission is dominated 
by a multitemperature, optically thin plasma component, 
at $kT \sim 0.2$--$0.7$ keV, 
slightly hotter in the nuclear region than in the arms 
(Figures 1, 2). It is likely to originate from gas 
shock-heated by core-collapse SN explosions\footnote{SN ejecta 
can easily provide 
$v_{\rm sh} \sim T^{1/2} > 650$ km s$^{-1}$, 
required to heat the gas to $kT \simgt 0.5$ keV. 
Assuming a SN rate $\sim 0.05$ yr$^{-1}$ for M\,83, 
and a total mechanical energy $\approx 10^{51}$ erg 
injected into the ISM by each SN, the total mechanical 
luminosity is $\approx 1.5 \times 10^{42}$ erg s$^{-1}$.}. 
The unresolved emission from the disk region 
has a power-law-like tail that dominates above 3 keV. 
Its origin is still unclear: a population of faint, 
unresolved XRBs can produce a power-law component. 
The emission can also be due to a second thermal plasma 
component at $kT > 2$ keV, or to Compton upscattering of far-IR photons 
by relativistic electrons (Valinia \& Marshall 1998).

From the observed temperature and luminosity, 
we estimate an average density $n_e \approx 5 \times 10^{-2}$ cm$^{-3}$ 
and a total mass $\sim 10^7$ $M_{\odot}$ for the X-ray emitting gas in M\,83; 
the cooling timescale is $\approx 10^8$ yr. For the hot gas 
in the starburst nuclear region, $n_e \approx 0.2$ cm$^{-3}$, 
$M \approx 5 \times 10^5$ $M_{\odot}$, $t_c \approx 4 \times 10^7$ yr.
Hence, the hot gas is an indicator of recent star formation.


\begin{figure}
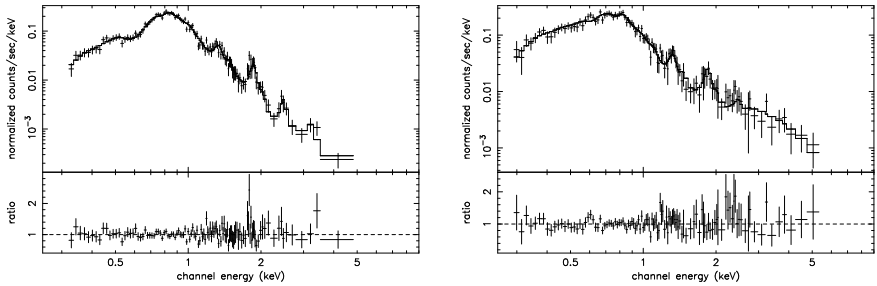
  
\vbox{
\begin{tabular}{lr}
\psfig{figure=diff16n2.ps,width=5.5cm, angle=270} &
\psfig{figure=armsnew1b.ps,width=5.5cm, angle=270} 
\end{tabular}
}
\caption{\footnotesize{{\it Left}: the {\it Chandra}/ACIS spectrum 
of the unresolved emission in the starburst nucleus of M\,83 shows 
a thermal plasma component 
($kT \approx 0.6$ keV, determined from the Fe L line 
complex) with strong metal lines; in particular, 
the 1.33 keV Mg\,XI line suggests that the gas 
has been enriched by core-collapse SNe.
{\it Right}: the unresolved emission in the spiral arms 
has a larger contribution from cooler gas ($kT \approx 0.2$ keV)
and a high-energy power-law component whose origin is yet to be determined.}}
\end{figure}


\section{Discrete X-ray sources}

\subsection{Population studies}

Populations of discrete X-ray sources in different galaxies and different 
galactic environments (bulge, disk, spiral arms, etc.) have different 
morphologies for their luminosity distributions (Fig.\ 3). 
It has been suggested that the slope and the break are indicative 
of the star-formation history (Wu 2001): 
an unbroken power-law indicates continuous star formation; 
a break (ie, a lack of bright sources) may be caused by aging 
of the X-ray source population, indicating the look-back time 
to the last major episode of star formation.
This in turn provides information on the dynamical history of a galaxy, 
because star formation is often triggered by close encounters and mergers with 
other galaxies.

Color-color plots can separate different classes 
of X-ray sources, and distiguish XRBs in the soft 
and hard state (Fig.\ 4). 
XRBs are, in turn, a mixture of young (timescale 
of $\sim 10^7$ yr after star formation), wind-accreting, 
high-mass XRBs (generally seen through higher intrinsic 
absorption), and old (timescale 
of $\sim 10^9$ yr after star formation), Roche-lobe accreting 
low-mass XRBs. For Galactic sources, the color 
and spectral separation between 
the hard and soft state is generally larger in XRBs with a BH accretor 
than in those with a NS.  

Studies of luminosity and color variability offer another criterion 
to separate XRBs (which often show state transitions) 
from SNRs (which do not). The age of different classes 
of X-ray sources can be inferred from a study of their 
spatial correlation with various indicators 
of recent star formation, eg, the spiral arms 
defined by the H\,{\footnotesize II} regions (Fig. 5).


\begin{figure}
\vspace{-1.2cm}
\vbox{
\begin{tabular}{lr}
\hspace{0.5cm}
\epsfig{file=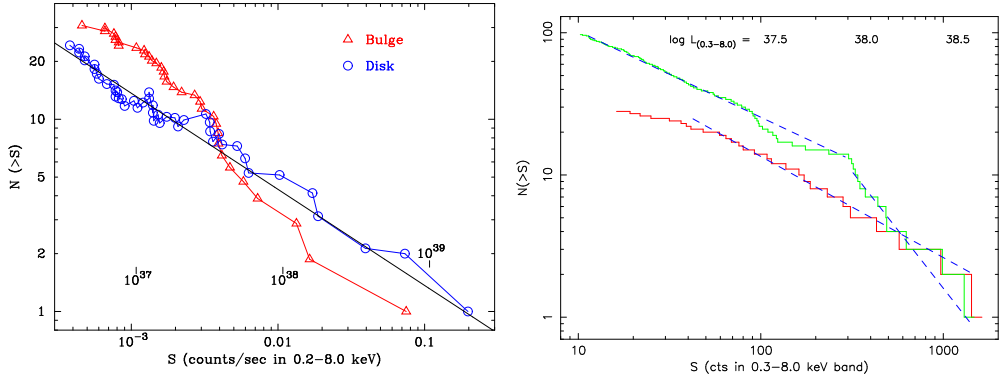,width=6.0cm} 
\hspace{-5.0cm}
\psfig{file=m83_lf.ps, width=4.8cm, angle=0}
\end{tabular}
}
\caption{\footnotesize{{\it Left}: the cumulative luminosity 
distribution of the discrete sources in M\,81 shows a break at
$L \approx 3 \times 10^{37}$ erg s$^{-1}$, suggesting 
that bulge sources belong to a much older population; 
continuous star formation in the disk 
results in an unbroken power-law instead (Tennant et al.\ 2001).
{\it Right}: in M\,83, the luminosity function of the sources 
inside the inner 60\arcsec~(nuclear starburst region + bar) does not 
show a break (lower curve). This can be interpreted as due to 
continuous star formation in the nuclear region. However, 
the luminosity function of the X-ray sources outside 60'' (upper curve)
has a break at $L_{\rm x} \approx 10^{38}$ erg s$^{-1}$,  
interpreted either as the upper limit of the NS population 
or as due to population aging (Soria \& Wu 2002).    
}} 
\end{figure}


\begin{figure}
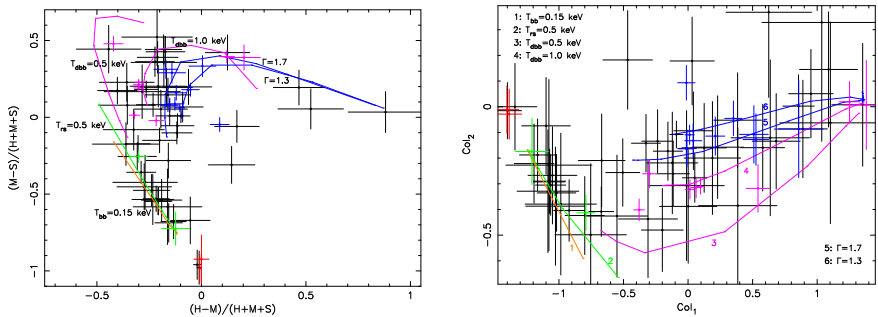
  
\vbox{
\begin{tabular}{lr}
\psfig{figure=m83_colplot1b.ps,width=5.5cm, angle=270} &
\psfig{figure=m83_colplot2b.ps,width=5.5cm, angle=270} 
\end{tabular}
}
\caption{{\footnotesize {\it Left}: The model lines 
in the X-ray color-color diagram constrain the expected 
locations (for $4 \times 10^{20} \leq n_{\rm H} \leq 
5 \times 10^{22}$ cm$^{-2}$) 
of different classes of X-ray sources in M\,83: 
XRBs in a hard state (power-law spectrum with 
$1.3 \simlt \Gamma \simlt 1.7$), XRBs in a soft state (diskbb with 
$0.5 \simlt kT \simlt 1.0$ keV), SNRs (Raymond-Smith thermal spectrum 
with $kT \approx 0.5$ keV), supersoft sources (blackbody spectrum 
with $kT \approx 0.06$ keV). 
Here $S = 0.3$--$1.0$ keV, $M = 1.0$--$2.0$ keV, $H = 2.0$--$8.0$ keV. 
For the brightest sources, we can obtain individual spectral fits. 
Datapoints of sources whose X-ray spectra are consistent 
with hard-state XRBs have been plotted in blue; plotted in magenta: 
soft-state XRBs; in green: SNRs; in red: supersoft sources.  
{\it Right}: Same model lines, 
for a different choice of X-ray colors. Here 
Col$_1$ $\equiv (CH +CM)/\sqrt{2}$, Col$_2$ $\equiv (CH -CM)/\sqrt{2}$, 
where $CH = (H-S)/(H+S)$, $CM = (M-S)/(M+S)$. 
}}
\end{figure}



\begin{figure}  
\vbox{
\begin{tabular}{lr}
\psfig{figure=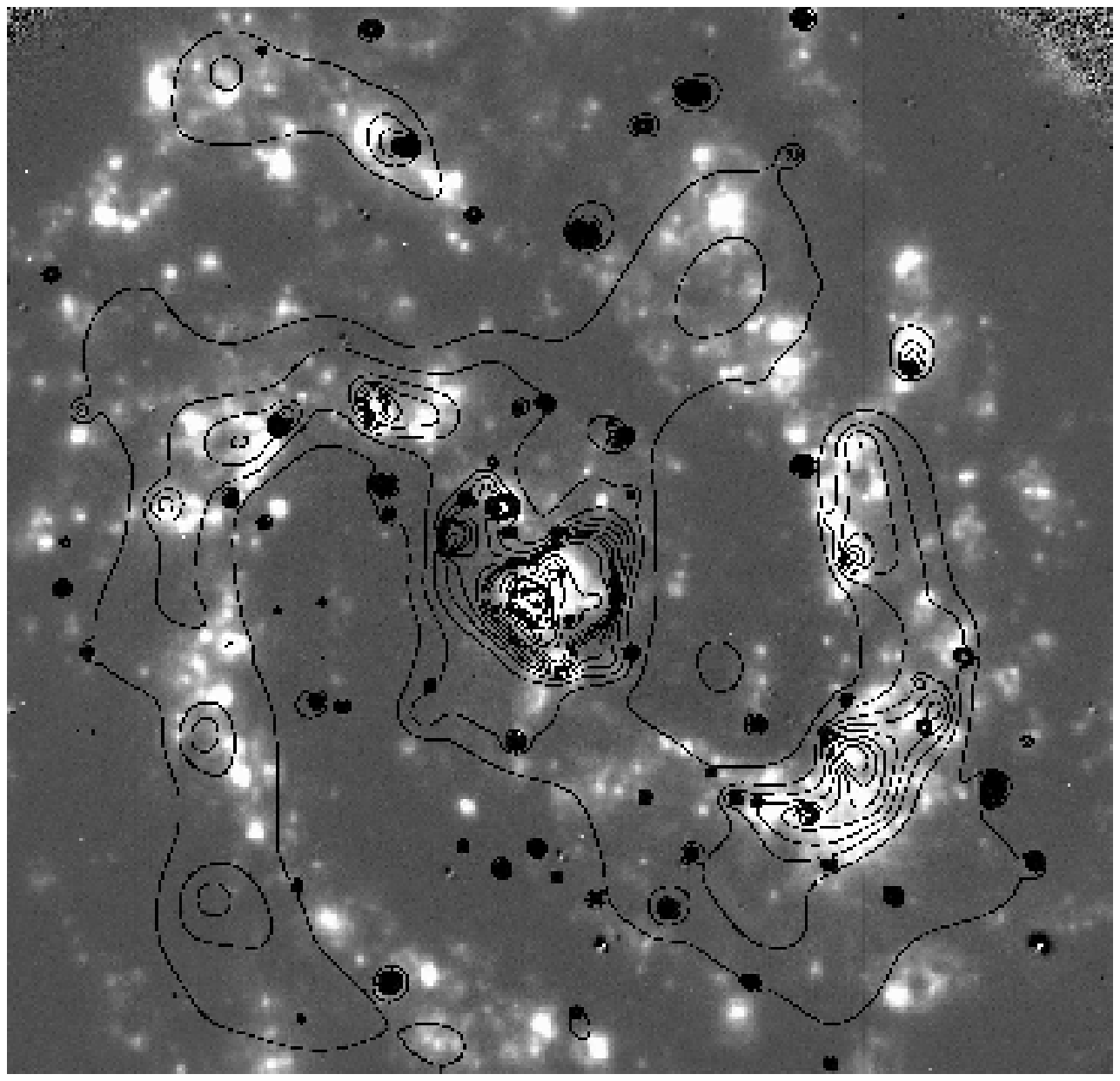,width=6.5cm} &
\psfig{figure=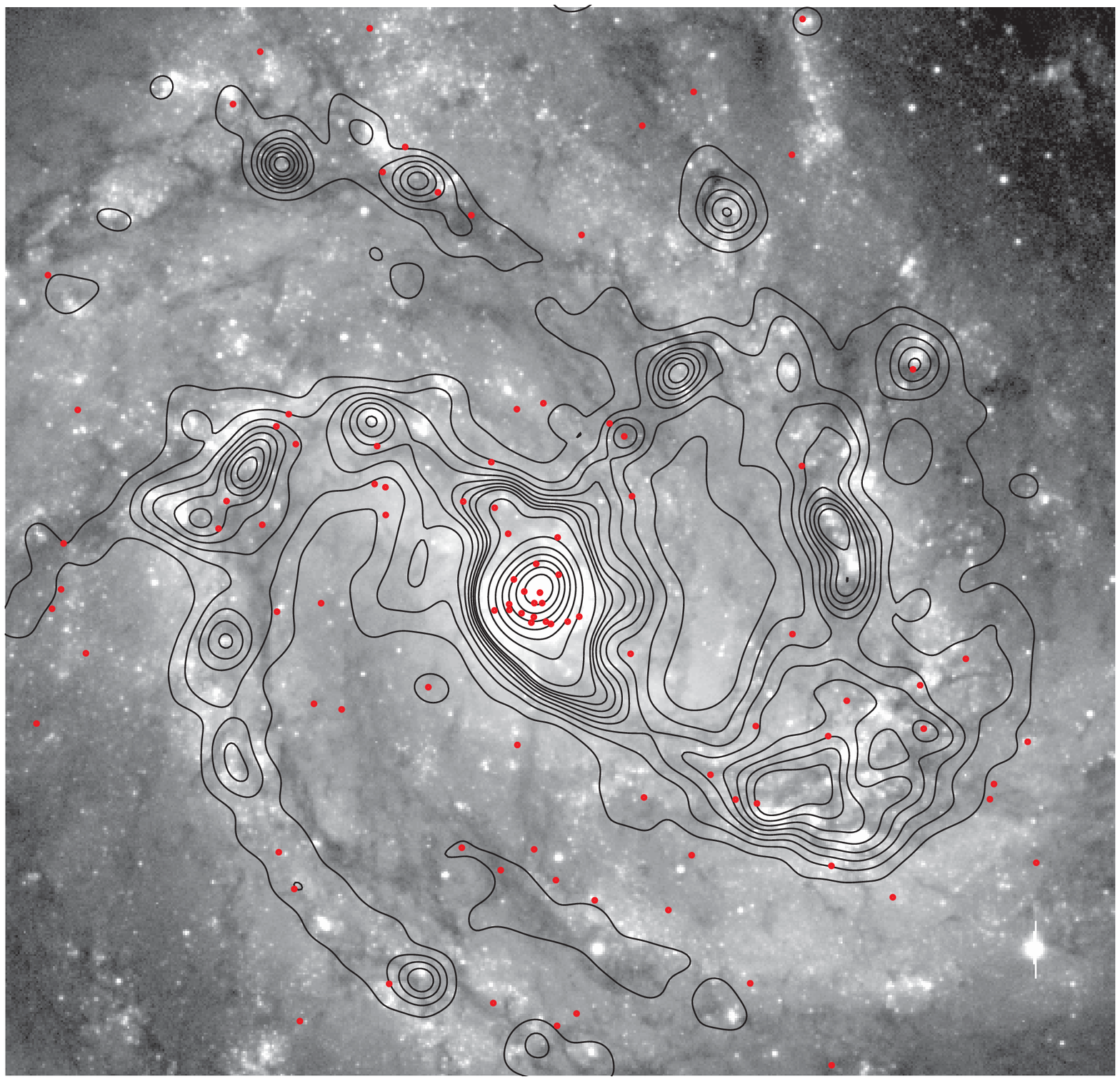,width=6.5cm} 
\end{tabular}
}
\caption{{\footnotesize {\it Left}: the diffuse X-ray 
emission (0.3--8.0 keV {\it Chandra} contours) 
is associated to the spiral arms and is an indicator 
of recent star formation, similar to the H$\alpha$ emission 
(greyscale image from the Anglo-Australian Telescope). 
{\it Right}: Most of the discrete X-ray sources (red circles) 
are associated with star-forming regions or young stellar populations 
(greyscale VLT B-band image, tracing OBA stars).
The radio flux (12 cm contours from a VLA image) 
is a combination of free-free and synchrotron emission 
and is also an indicator of star formation.
Size of both images: 6\arcmin $\times$ 6\arcmin. North is up, East is left.
}}
\end{figure}


\subsection{The nature of individual sources}

In M\,83, detailed spectral analysis is possible for sources 
with an emitted luminosity $\simgt 10^{38}$ erg s$^{-1}$. 
We can easily distinguish (Fig. 6):

\begin{itemize}
\item the X-ray nucleus, coincident with the optical/IR nucleus. 
Its spectrum is consistent with that of a supermassive black hole 
accreting well below its Eddington rate ($L_{\rm x} 
\approx 2 \times 10^{38}$ erg s$^{-1}$; $M \sim 10^7 M_{\odot}$);
\item XRBs in a hard state: their X-ray spectrum 
is well fitted with a simple power law of photon index $\approx 1.3$--$1.7$; 
\item XRBs in a soft state: their X-ray spectra 
is softer, dominated by a blackbody or disk-blackbody 
at $0.5 \simlt kT \simlt 1$, plus a power law of photon 
index $\approx 2.2$--$3.0$ (due to Compton upscattering of the disk photons);
\item ``emission-line'' sources: they show a few prominent metal lines 
(in particular from Si and Mg). The softer sources are more likely 
to be young SNRs; the harder ones could be X-ray binaries 
surrounded by a photoionized nebulae or stellar wind. 
They latter class would then be analogous to Vela X-1 or Cen X-3 
in our Galaxy (eg, Liedahl et al.\ 2000); 
\item bright supersoft sources, characterized by a thermal spectrum 
with blackbody temperature 
$kT \approx 60$ eV. They are generally interpreted as 
accreting white dwarfs with stable nuclear burning on their surface 
(see also Di Stefano \& Kong 2003).
\end{itemize}
For a more extensive discussion of the spectral 
and temporal properties of the discrete sources 
in M\,83, see Soria \& Wu (2003).

                             
\begin{figure*}
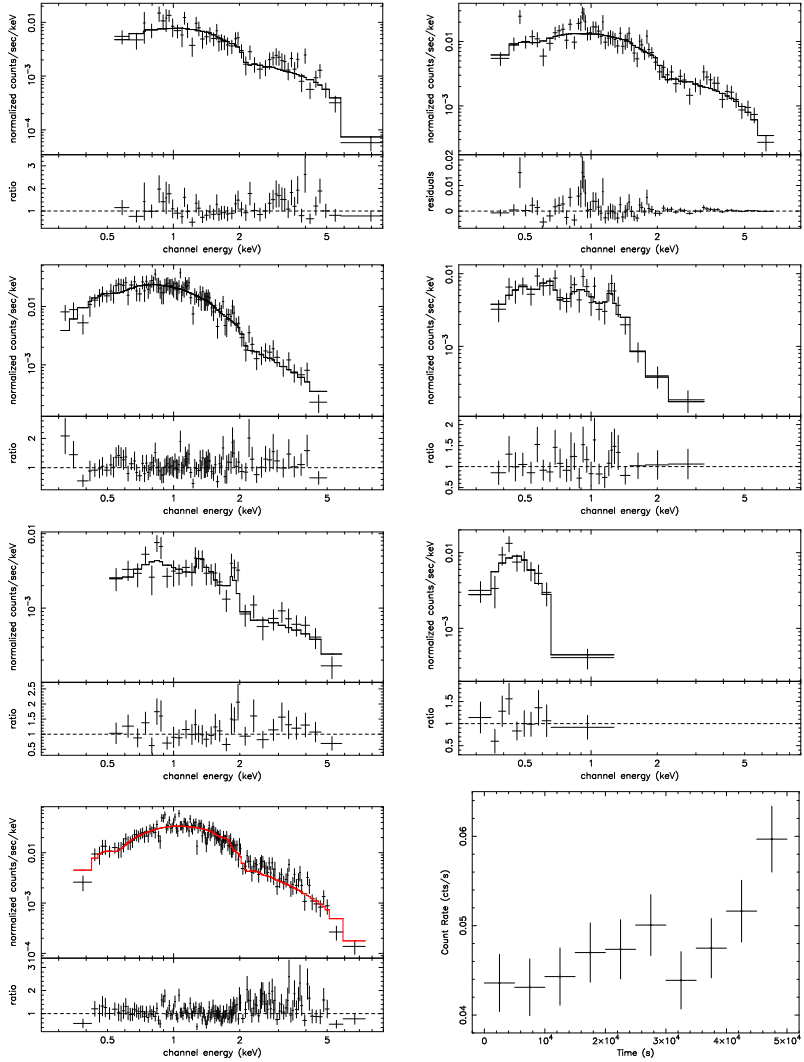

\vbox{
\begin{tabular}{lr}
\psfig{figure=s63.ps,width=5.0cm, angle=270} &
\psfig{figure=s44.ps,width=5.0cm, angle=270} \\
\psfig{figure=s64.ps,width=5.0cm, angle=270} &
\psfig{figure=snr08.ps,width=5.0cm, angle=270} \\
\psfig{figure=s27.ps,width=5.0cm, angle=270} &
\psfig{figure=ss1n.ps,width=5.0cm, angle=270} \\
\psfig{figure=ULX.ps,width=5.0cm, angle=270} &
\hspace{-0.4cm}
\psfig{figure=ulx_5000.ps,width=4.8cm, angle=270}
\end{tabular}
}
\caption{{\footnotesize Spectral fits to a sample of bright X-ray sources 
in M\,83 show different classes of objects. First three rows, from top left: 
the galactic nucleus; an X-ray binary in a hard state; an X-ray binary 
in a soft state; a source with soft thermal plasma emission (a young SNR?); 
a source with a power-law continuum plus line emission (from a photo-ionized 
stellar wind?); a supersoft source. See Soria \& Wu (2003) for further details 
on the sources and their fitted spectral models. Bottom row: 
the brightest X-ray source in M\,83 has an emitted luminosity 
in excess of $10^{39}$ erg s$^{-1}$. A power-law spectral fit 
gives a photon index $2.5 \pm 0.1$. Its lightcurve shows 
an increase by $\approx 40\%$ over the duration of the {\it Chandra} 
observation.
}}
\label{fig:image}
\end{figure*}


The brightest point source, located $\approx 5$\arcmin~south-east 
of the nucleus, is variable and 
has an emitted luminosity of $\approx 10^{39}$ 
erg s$^{-1}$ (Fig. 6, bottom panels). 
Hence, it can be classified as an ultra-luminous source (ULX).

\section{Ultra-luminous sources}

\subsection{X-ray observations}

The maximum luminosity of an accreting source is known as 
Eddington limit:
\begin{displaymath}
L_{\rm Edd} = \frac{4\pi c G M}{\langle \kappa \rangle } 
  = 1.3 \times 10^{38} \, 
  \left(\frac{\kappa_{\rm Th}}{\langle \kappa \rangle }\right)\, 
  \left(\frac{M}{M_{\odot}}\right)\ {\rm erg\ s}^{-1}.
\end{displaymath}
where $\langle \kappa \rangle$ is the average flux-weighted opacity.

The term ``ULX'' is generally applied to X-ray sources 
outside a galactic nucleus which persistently exceed  
(when in an active state) the Eddington 
luminosity of a 7-$M_{\odot}$ BH (thought to be the ``canonical'' 
BH mass value); some sources emit a luminosity 
of up to a few times $10^{40}$ erg s$^{-1}$ (Roberts \& Warwick 2000). 
In fact, it is necessary 
to distinguish between bright SNR (for which the Eddington limit 
does not apply) and accreting sources. Even in the absence 
of detailed spectral information, variability 
by a factor $\simgt 2$ is usually taken as a good indicator  
that a source is not an SNR. Correlation 
with a non-thermal radio source is another criterion 
to identify an SNR. 

It is still unclear whether ultra-luminous accreting sources are simply 
the high-luminosity end of the X-ray binary population, 
or a different physical class of objects. 
No significant break or feature at $L \approx 10^{39}$ erg s$^{-1}$ 
is found in the cumulative luminosity distribution 
of X-ray sources in galaxies which contain ULXs. However, 
the statistical error is large, due to the small number 
of sources in each galaxy above that luminosity.

ULXs are found in elliptical galaxies (eg, NGC 1553: Sarazin et al. 2000; 
NGC 4697: Blanton, Sarazin \& Irwin 2001), associated with 
an old stellar population, often inside globuler clusters. 
Their location, and the steep slope of the high-luminosity 
end of the X-ray luminosity function suggest that, in this case, 
they are old systems accreting from a low-mass companion.  
On the other hand, ULXs are also often found 
in starburst or active star-forming galaxies 
(eg, M82: Matsumoto et al. 2001; the Antennae: Zezas \& Fabbiano 2002). 
The X-ray luminosity function in these galaxies
is typically an unbroken power-law with a flat slope, 
dominated at high luminosities by young high-mass XRBs.   
In this case, the ULXs appear to be associated with 
a very young stellar population.

X-ray spectral analyses do not provide a unique physical 
identification, either. Some ULXs can be fitted with a simple 
power law continuum (La Parola et al. 2001; Strickland et al. 2001).  
Others are better fitted with a disk-blackbody model 
(Makishima et al. 2000; Roberts et al. 2002) with color 
temperatures $\approx 1$--1.5 keV (such high color temperatures 
can still be consistent with a high-mass accretor, if the spectral 
hardening factor is $\approx 3$).
Transitions from a hard to a soft state are seen in a few cases 
(Kubota et al. 2001), analogous to those detected 
in Galactic BH candidates. At least one ULX has a supersoft spectrum with 
a blackbody temperature of $\approx 80$ eV (Swartz et al. 2002).
Finally, X-ray variability studies have shown 
a range of different behaviours: most of the sources are persistent;  
a few are transients on timescales of a few months/years;  
others are highly variable on timescales of a few thousand seconds 
(a ULX in M74: M. Garcia 2002, priv. comm.)

\subsection{Three physical models for the ULXs}

There are at least three models to explain the physical nature 
of the ULXs (having excluded the bright X-ray SNRs from this category):

\begin{itemize}
\item {\bf{Intermediate-mass black holes (IMBHs)}}.\\
 The Eddington limit is proportional to the mass of the accreting 
object, hence a 100-$M_{\odot}$ BH could have a persistent 
luminosity $\simgt 10^{40}$ erg s$^{-1}$. The first problem 
this model has to address is how to form them. 
It is still not known what the maximum mass is for BHs formed 
via SN explosions of single stars, and how this mass depends 
on the metallicity of the progenitor star. If BH masses $\simgt 50 M_{\odot}$ 
are required to explain the observations, mergers 
of smaller-scale bodies in a dense environment 
are likely to be necessary.  
It has been shown (Sigurdsson \& Hernquist 1993; 
Kulkarni, Hut, \& McMillan 1993) that it is not possible to merge 
small ($M \simlt 50 M_{\odot}$) BHs in a globular cluster via three body 
interactions: recoils tend to kick the BHs out of the cluster 
before they can merge. However, mergers become possible 
if one considers four-body interactions (Miller \& Hamilton 2002). 
The best chance to form an IMBH occurs if a number of 
progenitor stars coalesce first into a $\sim 10^3 M_{\odot}$ star 
which would then undergo a SN explosion (Ebisuzaki et al. 2001). 
In this case, the formation process would be 
most likely to occur near the center of compact, young super-star clusters 
usually found in starburst galaxies. If super-star clusters 
are the progenitors of globular clusters, this could also 
explain the presence of IMBHs in the old globulars of elliptical galaxies.
Yet another suggestion (Madau \& Rees 2001) 
is that IMBHs could be pre-galactic remnants 
formed from the fragmentation of primordial molecular clouds.
The second problem to address is how to feed them: a discussion 
on the mass transfer mechanism (Roche-lobe overflow 
or stellar wind) and allowed ranges of mass for the companion star is beyond 
the scope of this review. (See Zezas \& Fabbiano 2002 for such 
a discussion regarding the ULXs in M\,82).  
Thirdly, how to observe them. The most reliable determination 
of the mass function for Galactic BHs comes from optical observations 
of photometric variations and line velocity shifts 
of the accretion disk and companion star over a binary period. 
It is of course more difficult 
to achieve that in other galaxies (for M\,83, distance modulus $\approx 28$).

\item {\bf{Beating the Eddington limit}}\\
The classical Eddington limit assumes that the average opacity 
$\langle \kappa \rangle \geq \kappa_{\rm Th}$. 
However, this may not be the case if the radiating medium 
(accretion disk or stellar surface) is clumpy. It can be shown 
that the flux-weighted, volume-averaged opacity can be lower 
than the electron scattering opacity for an inhomogeneous medium 
(Shaviv 1998). If this is the case, more photons can escape without 
blowing away the accretion flow, 
so that steady-state 
luminosities $\sim 10^{40}$ erg s$^{-1}$ could in principle 
be attained by a stellar-mass BH with $M \simlt 10 M_{\odot}$.
Observational evidence for ``super-Eddington'' luminosities 
has been suggested in the case of novae (Shaviv 2001). Models 
of clumpy accretion disks based on the same principle have 
also been proposed (Begelman 2002). 
At $L \approx L_{\rm Edd}$, all systems 
should develop strong radiation-driven winds, whose density and 
velocity could in principle be inferred from high-resolution 
UV and X-ray spectra.

\item {\bf{Beamed emitters}}\\
If the X-ray emission is beamed towards us rather than isotropic, 
the estimated ULX luminosities can be scaled down, 
depending on the beaming factor. Beamed X-ray emission 
is known to occur in some microquasars (eg, GRS 1915$+$105). 
Such systems would appear highly super-Eddington if we happen 
to observe them pole-on (Fabrika \& Mescheryakov 2001; King et al. 2001).
It is sometimes possible to determine whether the X-ray emission 
is beamed by studying the optical spectral lines of the photoionized 
nebula around a ULX.  A detailed discussion of this technique 
is presented by M. Pakull elsewhere in these Proceedings. 

\end{itemize}

\vspace{0.2cm} 

\noindent
{\bf Acknowledgements}\\
Thanks to Kinwah Wu, Rosanne Di Stefano, 
Albert Kong, Manfred Pakull, Andrea Prestwich, Doug Swartz 
for comments and discussions, to Stuart Ryder for the use 
of his AAT images, and to Reiner Beck for his VLA radio map.

\end{document}